\documentclass[sn-mathphys-num, iicol]{sn-jnl}
\usepackage{graphicx}%
\usepackage{multirow}%
\usepackage{amsmath,amssymb,amsfonts}%
\usepackage{amsthm}%
\usepackage{mathrsfs}%
\usepackage[title]{appendix}%
\usepackage{xcolor}%
\usepackage{textcomp}%
\usepackage{manyfoot}%
\usepackage{booktabs}%
\usepackage{algorithm}%
\usepackage{algorithmicx}%
\usepackage{algpseudocode}%
\usepackage{listings}%
\usepackage{subfigure}
\usepackage{orcidlink}

\begin{document}

\title[Measuring the ATLAS ITk Pixel Detector Material via Multiple Scattering of Positrons at the CERN PS]{Measuring the ATLAS ITk Pixel Detector Material via Multiple Scattering of Positrons at the CERN PS}

\author*[1]{\fnm{Simon Florian} \sur{Koch}\,\orcidlink{0000-0002-2676-2842}}\email{simon.florian.koch@cern.ch}

\author*[1,2]{\fnm{Brian} \sur{Moser}\,\orcidlink{0000-0001-6750-5060}}\email{brian.moser@cern.ch}

\author[2]{\fnm{Anton\'in} \sur{Lindner}\,\orcidlink{0009-0002-5800-3872}}

\author[2,5]{\fnm{Valerio} \sur{Dao}\,\orcidlink{0000-0003-1645-8393}}

\author[2]{\fnm{Ignacio} \sur{Asensi}\,\orcidlink{0000-0002-0332-9229}}

\author[1]{\fnm{Daniela} \sur{Bortoletto}\,\orcidlink{0000-0002-1287-4712}}

\author[2]{\fnm{Marianne} \sur{Brekkum}}

\author[2]{\fnm{Florian} \sur{Dachs}}

\author[2,3]{\fnm{Hans Ludwig} \sur{Joos}\,\orcidlink{0000-0003-4313-4255}}

\author[2,4]{\fnm{Milou} \sur{van Rijnbach}\,\orcidlink{0000-0003-3728-5102}}

\author[2]{\fnm{Abhishek} \sur{Sharma}\,\orcidlink{0000-0002-5211-7177}}

\author[2]{\fnm{Ismet} \sur{Siral}\,\orcidlink{0000-0003-4554-1831}}

\author[2]{\fnm{Carlos} \sur{Solans}\,\orcidlink{0000-0002-0518-4086}}

\author[1]{\fnm{Yingjie} \sur{Wei}\,\orcidlink{0000-0001-9725-2316}}


\affil[1]{\orgdiv{Department of Physics}, \orgname{Oxford University}, \orgaddress{\city{Oxford}, \country{UK}}}

\affil[2]{\orgname{CERN}, \city{Geneva}, \country{Switzerland}}

\affil[3]{\orgdiv{II. Physikalisches Institut}, \orgname{Georg-August-Universität G\"ottingen}, \orgaddress{\city{G\"ottingen}, \country{Germany}}}

\affil[4]{\orgdiv{Department of Physics}, \orgname{University of Oslo}, \orgaddress{\city{Oslo}, \country{Norway}}}

\affil[5]{\orgdiv{Departments of Physics and Astronomy}, \orgname{Stony Brook University}, \orgaddress{\city{Stony Brook}, \state{NY}, \country{USA}}}


\abstract{\unboldmath The ITk is a new silicon tracker for the ATLAS experiment designed to increase detector resolution, readout capacity, and radiation hardness, in preparation for the larger number of simultaneous proton-proton interactions at the High Luminosity LHC. This paper presents the first direct measurement of the material budget of an ATLAS ITk pixel module, performed at a testbeam at the CERN Proton Synchrotron via the multiple scattering of low energy positrons within the module volume. Using a four plane telescope of thin monolithic pixel detectors from the MALTA collaboration, scattering datasets were recorded at a beam energy of $1.2\,\text{GeV}$. Kink angle distributions were extracted from tracks derived with and without information from the ITk pixel module, and were fit to extract the RMS scattering angle, which was converted to a fractional radiation length $x/X_0$. The average $x/X_0$ across the module was measured as $[0.89 \pm 0.01 \text{ (resolution)} \pm 0.01 \text{ (subtraction)} \pm 0.08 \text{~(beam momentum band)}]\%$, which agrees within uncertainties with an estimate of $0.88\%$ derived from material component expectations.}

\maketitle

\section{Introduction}
\label{sec:intro}

The High Luminosity upgrade of the LHC (HL-LHC) will increase the instantaneous luminosity of the machine up to $7.5 \times 10^{34}\,\text{cm}^{-2}\,\text{s}^{-1}$, facilitating an order of magnitude increase in the delivered proton-proton collision data~\cite{ZurbanoFernandez:2020cco}. This increase in instantaneous luminosity will proportionally increase data rates and radiation damage in the ATLAS detector. The entire inner detector system will be replaced by an all-silicon tracker, the ITk~\cite{CERN-LHCC-2017-005, CERN-LHCC-2017-021, ATL-PHYS-PUB-2019-014}, to keep excellent physics performance in these harsh environments. The ITk will improve over the currently installed tracker by having a higher granularity, an improved radiation hardness and an increased coverage in the forward direction. Requirements on the tracking and vertexing performance of the ITk mandate the detector to have a lower material budget, especially for the innermost layers, which will feature silicon pixel detectors. Accurate knowledge of this material content is paramount to ensure the correctness of performance estimations and in turn the expected physics output of the experiment. It is also a key incredient to the simulations that will later on be compared to data in the measurements themselves. This paper presents a first measurement of the fractional radiation length of an ITk pixel module via the multiple scattering of $1.2\,\text{GeV}$ positrons at the T9 beamline of the CERN Proton Synchrotron. Alongside the measurement methodology, the design and construction of the purpose-built MONSTAR beam telescope will be discussed, as will the analysis method and results.

\begin{figure}[h!]
  \centering
  \includegraphics[width=0.45\textwidth]{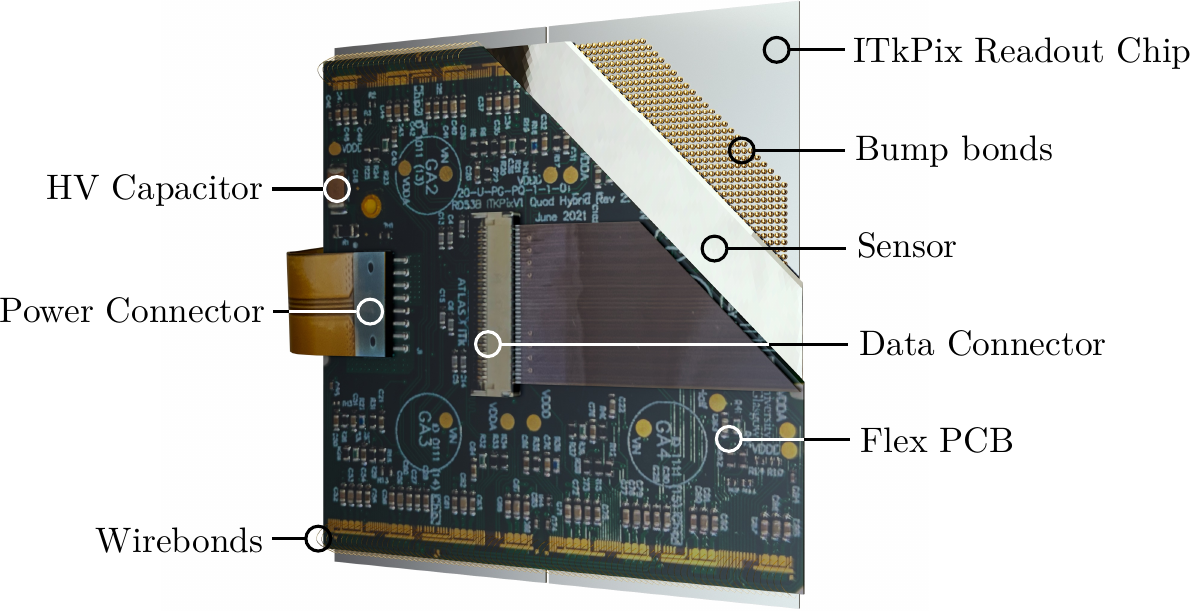}
  \caption{Annotated render of an ITkPix quad module, with a cutaway revealing the main layers. The high voltage (HV) capacitor and the data and power connectors have been highlighted as these are substantial contributions to the material budget of the module.}
  \label{fig:module}
\end{figure}

The pixel modules used in the ITk vary in sensor technology depending on their position in the detector. The innermost layer uses 3D sensors, while the outer layers use planar sensors with varying thickness. A hybrid pixel module design is used, with sensors bump bonded to readout chips (ROCs), which are in turn wire bonded to a printed circuit board (PCB) with connectors to the outside world. The ROCs were developed by the RD53 collaboration~\cite{Garcia-Sciveres:2665301} and varying pixel pitches are used within the sensor layout, depending again on the location of the module in the detector. The module used in this measurement is a quad module designed for the Outer Barrel part of the ITk pixel detector, consisting of 4 ROCs in a $2 \times 2$ matrix bump bonded to a single silicon sensor. A cutaway view of the module is shown in Figure \ref{fig:module}. Especially the populated PCB with its connectors, pads and surface mount devices (SMDs) makes the module an ideal target for a spatially resolved fractional radiation length measurement. This is further exacerbated by the fact that for many of these components it is difficult to get an accurate estimate of their radiation length.

Multiple (Coulomb) scattering can be described within Moli\`ere theory~\cite{PhysRev.89.1256}, predicting the scattering angle distribution to be composed of three analytic terms: a Gaussian core analogous to the central limit of a large number of soft scatters, a transition to the Rutherford formula for large angle scatters, and a correction. The root mean square (RMS) of the Gaussian core of the scatter angle projected onto a reference plane parallel to the incident direction can be related to the fractional radiation length $x/X_{0}$ with the Lynch \& Dahl revision of the Highland formula\footnote{In the following referred to as the Highland formula for the sake of readability.}~\cite{Lynch:1990sq, highlandpdg} as

\begin{align}
\begin{split}
&\theta^{\text{RMS}}_{\text{Highland}} = \\ &\frac{13.6\,\text{MeV}}{\beta c p} z \sqrt{x/X_{0}}\left[1 + 0.038\ln\left(\frac{xz^2}{X_{0}\beta^2}\right)\right] \quad \text{.}
\label{eq:highland} 
\end{split}
\end{align}

Here, $\beta c$, $p$ and $z$ are the speed, momentum, and charge number of the particle that undergoes multiple scattering. This formula can be analytically inverted to calculate the fractional radiation length from the measured distribution of scatter angles. An alternative description by Fr\"uhwirth and Regler~\cite{FRUHWIRTH2001369}, which derives from a fit to a large convolution of single scattering distributions, improves over the Highland formalism by adding one additional parameter in the fit to derive the Gaussian core expression.

Multiple scattering based radiation length measurements at testbeams have been performed before, for example using a six plane MIMOSA telescope~\cite{STOLZENBERG2017173, Poley:2021wcw, Qu:2021qda}, but generally focussed on relatively thick subjects in relation to a pixel module, and did not attempt to image instrumented detectors. A recent measurement by the Tracker Group of the CMS Collaboration imaged an upgrade pixel module at the PSI PiE1 beamline~\cite{TrackerGroupoftheCMS:2024utw}. This work builds on the previous to design and build a four plane telescope based on MALTA detectors~\cite{SolansSanchez:2023pik}, and perform the first measurement of the radiation length of an entire ITk pixel module with sub-mm resolution and $\mathcal{O}(10\%)$ uncertainties. We present results for the fractional radiation length obtained with Highland's formalism and the improved Fr\"uhwirth-Regler description, as well as from a parametric fit to GEANT4 based simulation.

\section{Experimental setup}
\label{sec:setup}

\begin{figure}[h!]
  \centering
  \includegraphics[width=0.45\textwidth]{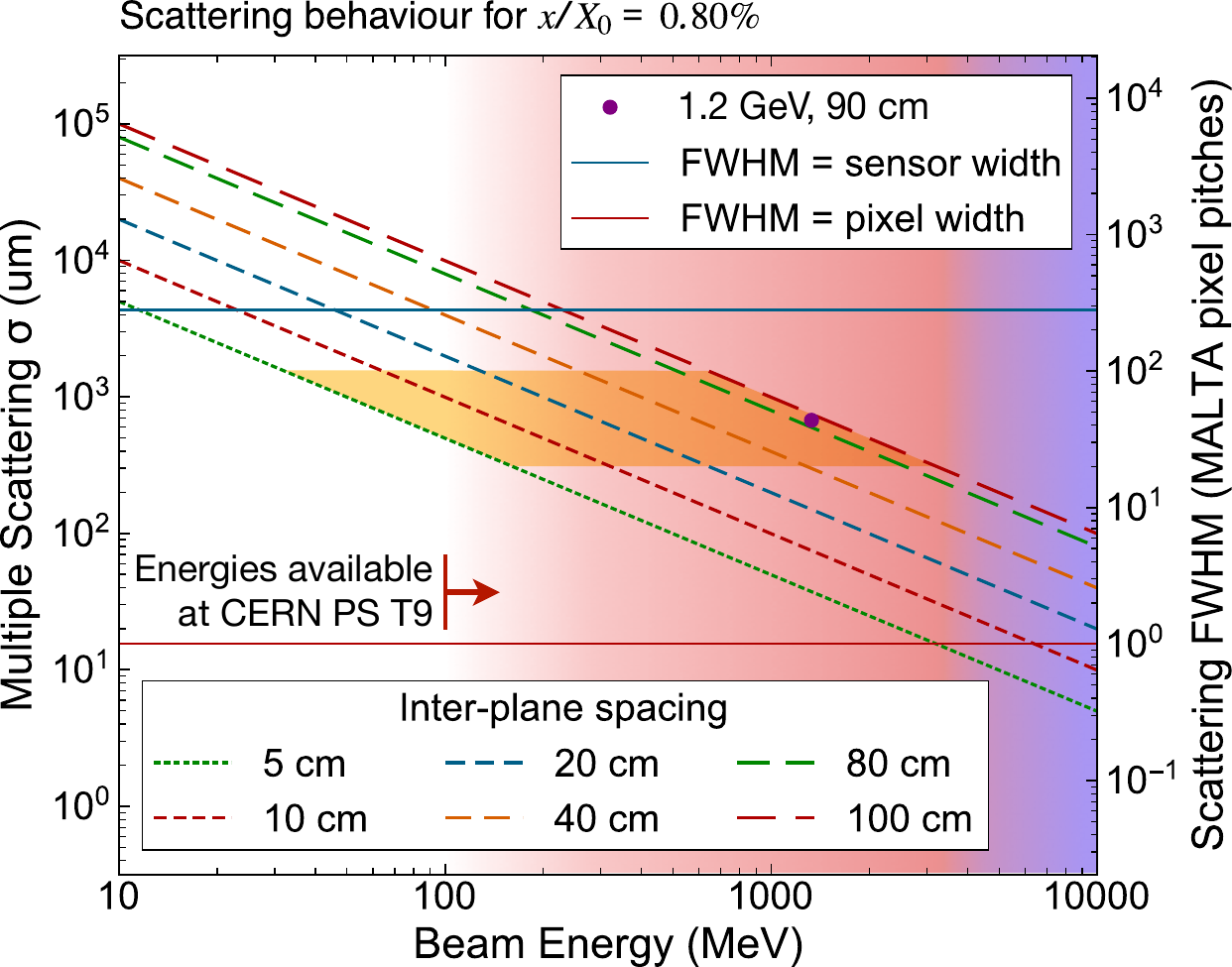}
  \caption{Relationship between beam energy and multiple scattering distribution width predicted for electrons/positrons for varying spacing distances between a subject with $x/X_0 = 0.8\%$ and the furthest downstream plane. The RMS of the scatter distribution $\sigma$, which equals $\theta^{\text{RMS}}$ in Equation \ref{eq:highland}, is shown in units of $\mu \text{m}$ (left) and the FWHM of the distribution is shown in units of Malta pixel pitches (right). The red and blue color backgrounds indicate the energy range where the PS T9 beam is dominated by positrons and pions, respectively. The intensity of the color is indicative for the expected rate. The horizontal blue line marks the upper limit where the scatter FWHM equals the sensor width, the horizontal red line marks the lower limit where the scatter FWHM equals the pixel pitch. The area highlighted in yellow indicates the expected telescope geometry and beam energy constraints for such measurements. The measurement presented in this paper is indicated as a violet point.}
  \label{fig:scattering}
\end{figure}

Using Eq.~\ref{eq:highland}, a relation between the beam energy and the width of the scatter angle distribution for a given target $x/X_{0}$ and inter-plane spacing can be derived. This relationship is illustrated in Figure~\ref{fig:scattering} using both the standard deviation expressed in $\mu\text{m}$ on the furthest downstream plane as well as the FWHM in units of pixel pitches of the telescope reference planes. For this comparison, the fractional radiation length of the device under test (DUT) was assumed to be $0.8\%$, corresponding to early estimates for the pixel module under test. The plot maps out the phase space to be considered for the measurement.

Non-hadronic beams are preferred for this measurement, to minimise the amount of hadronic interactions. Out of all beams that the T9 beam line at the CERN PS can provide, a $1.2\,$GeV positron beam was chosen for the main data-taking as it represented a compromise between purity, rate, and not requiring an unreasonably long telescope.

\begin{figure*}[h]
  \centering
  \includegraphics[width=0.65\textwidth]{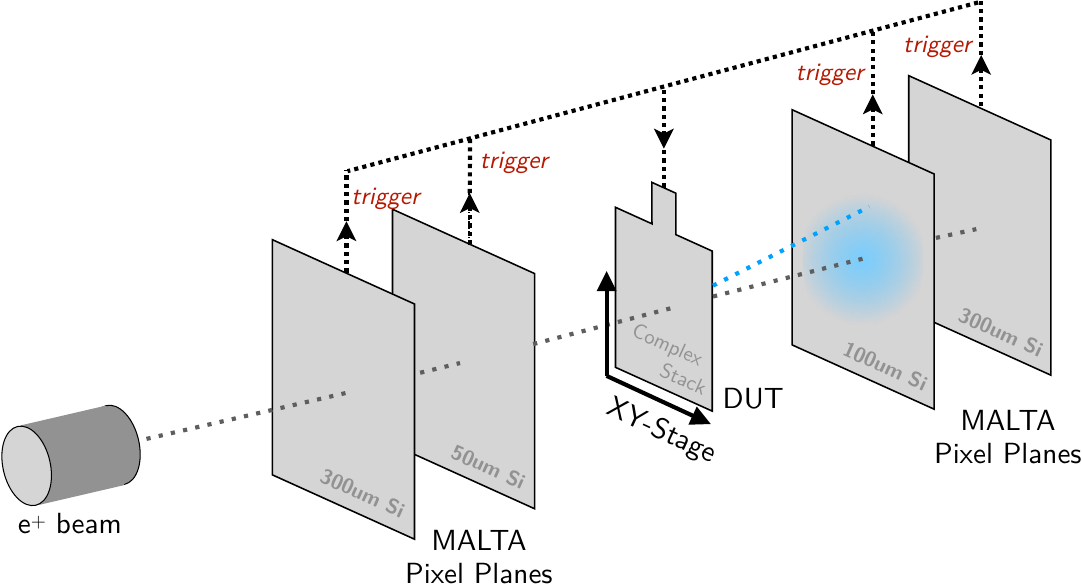}
  \caption{MONSTAR telescope configuration and trigger scheme. Multiple scattering within the DUT produces track kinks, which are analysed to derive the radiation length of the DUT.}%
  \label{fig:telescope_sketch}
\end{figure*}

\begin{figure*}[h]
  \centering
  \includegraphics[width=0.85\textwidth]{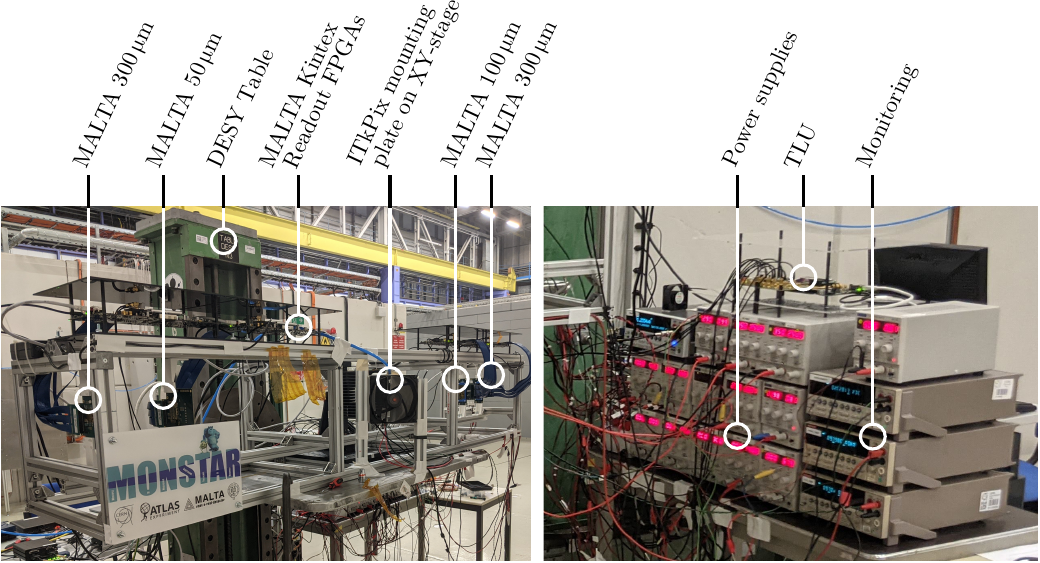}
  \caption{MONSTAR telescope setup during the CERN PS testbeam measurement.}%
  \label{fig:telescope_picture}
\end{figure*}

Following this choice, the MONSTAR telescope (Multiple Or Negligible Scattering Telescope, As Required) was constructed. It is a 4 plane telescope allowing the installation of a DUT on a $x-y$ linear stage as the central plane. The telescope frame is large enough to accomodate inter-plane spacing distances from $\sim 5\,$cm to $\sim 80\,$cm. The telescope setup is shown in Figure~\ref{fig:telescope_sketch}. The reference planes contain depleted monolithic active pixel sensors from the MALTA collaboration, with $512 \times 512$ pixels each and a pixel pitch of $36.4 \times 36.4~\mu$m$^2$. Two $300\,\mu$m thick Czochralski-type sensors~\cite{Pernegger:2023yaf} are used as outer planes, since they provide larger cluster sizes, and hence improved spatial resolution through use of a centre of gravity clustering algorithm. The inner two planes are $50\,\mu$m (upstream) and $100\,\mu$m (downstream) thick and contain less material to minimise the amount of multiple scattering within the telescope. Placed close to the DUT, these planes allow for long lever arm tracking without requiring data from the DUT itself, potentially allowing the telescope to be used for radiation length measurements on non-instrumented subjects. They are furthermore included in the trigger decision to reduce the rate of fake (noise) triggers. Trigger signals are generated from the coincidence of all four MALTA planes. An annotated photograph of the telescope is shown in Figure~\ref{fig:telescope_picture}. The readout system for the reference planes and the trigger logic unit (TLU) are loaded on Xilinx Kintex development boards and have been adopted from past MALTA beam telescopes~\cite{vanRijnbach:2023qgn}.

\begin{figure*}[h]
  \centering
  \hfill
  \subfigure[Photograph of installation in telescope.]{\includegraphics[width=0.45\textwidth]{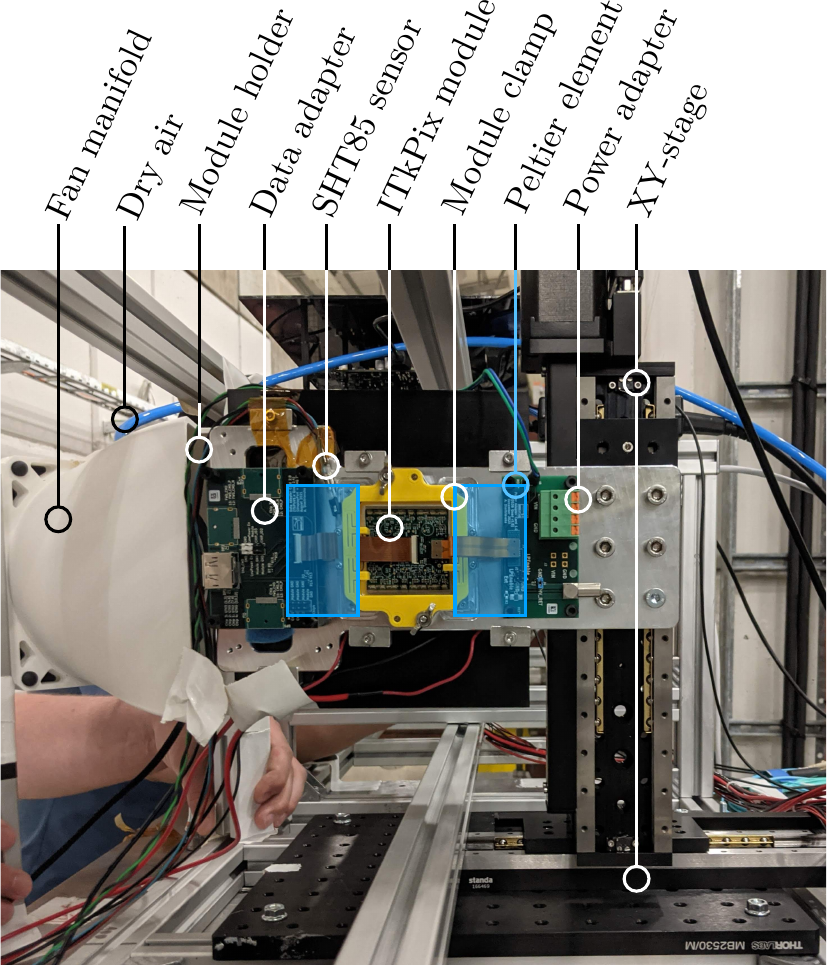}}
  \hfill
  \subfigure[Design render.]{\includegraphics[width=0.35\textwidth]{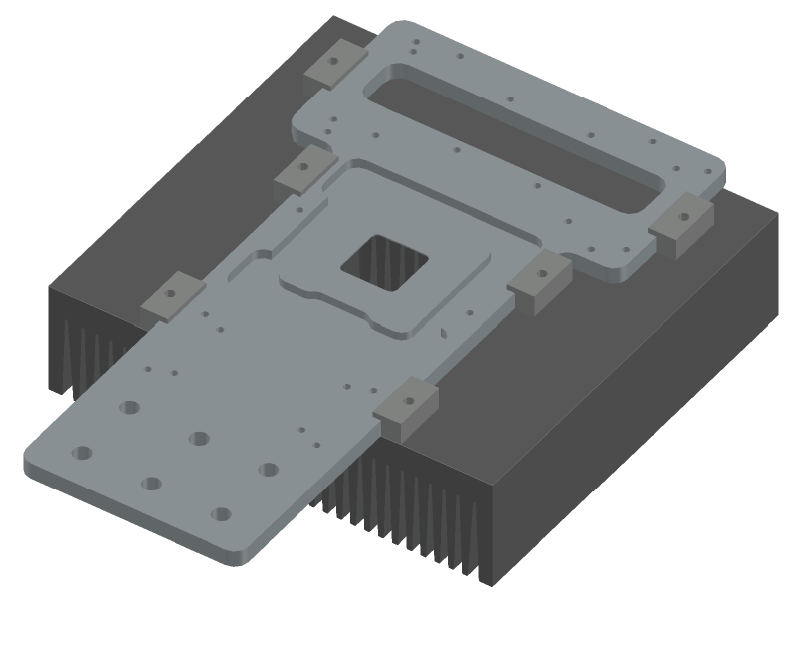}}
  \hfill
  \caption{Custom mounting plate used to provide mechanics and cooling for the ITkPix module, annotated with the relevant components. The location of the Peltier elements between the holder and heatsink is denoted by the blue shaded regions. The window size beneath the module used during the measurement was $35 \times 25\,\text{mm}^2$.}
  \label{fig:module_holder}
\end{figure*}

A pair of linear stages provide horizontal and vertical positioning capability of the DUT within the beam. A dedicated mounting plate was used to hold the ITk pixel module and provide cooling through the usage of two Peltier elements in combination with a heat sink and an air convection fan. The setup is shown in Figure \ref{fig:module_holder}. A rectangular window in the mounting plate below the sensor ensures that the plate does not contribute to the material budget. Only the a small fraction at the edges of the sensor is in physical contact with $450\,\mu$m thick aluminium to allow for sufficient heat transport to prevent the module from overheating\footnote{The exact dimensions of the widow were optimised using finite element analysis simulations for the heat transport.}. The physical contact is obtained through a 3D printed clamp that pushes the module on three edge points onto the aluminium. During operation, this cooling technique allowed to keep the module temperature constantly below 40$^{\circ}\text{C}$.

The ITk pixel module readout follows the same setup as the one foreseen in the actual detector. Therefore, this measurement also constitutes a readout test under realistic data taking conditions with external triggering. The signal is first transmitted via electrical links to an Optoboard~\cite{Franconi_2022} where it is converted to an optical signal fed into a readout server using FELIX~\cite{Ryu_2017, Solans_Sanchez_2020}. The readout software was integrated into the MALTA telescope sortware framework to allow for real time hit monitoring during data taking. Trigger input was accepted directly into the FELIX system as an LVDS signal from the MALTA TLU.

Data was taken using $1.2\,\text{GeV}$ positrons with a 5\% momentum band, imaging the full $4 \times 4\,\text{cm}^2$ region of the ITk pixel module in 16 steps of $1 \times 1\,\text{cm}^2$. The inner planes were moved close to the DUT to allow for tracking also without DUT hit information. At least 2 million triggers were collected in each step. In addition, a reference dataset was taken with a $50\,\mu$m MALTA detector as DUT.

\section{Analysis}
\label{sec:analysis}

A combination of the MALTA telescope software and the Corryvreckan testbeam analysis suite~\cite{Dannheim_2021} is used for data quality checks and alignment, as well as tracking. A global frame is defined with the $z$ axis pointing in beam direction and $x$ and $y$ chosen to align with the column and row directions of the DUT.

Noisy pixels are masked in each run separately using a local density method. If the pixel occupancy exceeds the local average by more than five standard deviations, it is masked and hits in this pixel are disregarded. A correction is added to the method to account for the higher occupancy of the central four rows and columns, which are double length to cover the inter-chip region on the quad module, and also exhibit high crosstalk. Trigger synchronisation is checked following a method described in Ref.~\cite{TrackerGroupoftheCMS:2024utw} which relies on correlation of the $x$ (and $y$) coordinates of hits across the different telescope planes in the synchronised case. No loss of synchronisation is observed for the data that was taken. Alignment is performed in two stages for each run, starting with a correlation-based pre-alignment followed by the application of the Millepede-II algorithm~\cite{Blobel:2002ax} to perform residual minimisation in $x$, $y$ and $\theta_z$ (the plane rotation angle about the $z$-axis) for each plane, with all other degrees of freedom fixed.

Multiplet tracking is used to determine the incident and outgoing vectors from which the scatter angle is calculated. Separate straight line fits are performed to hits upstream and downstream of the DUT, with the DUT hit included if not specified otherwise. Accepted tracks are required to include one hit in any active plane. Uncertainties on the track origin and slope are computed from the linear fit, and propagated to the extracted angles between the track segments.

\begin{figure*}[t!]
  \centering
  \includegraphics[width=0.65\textwidth]{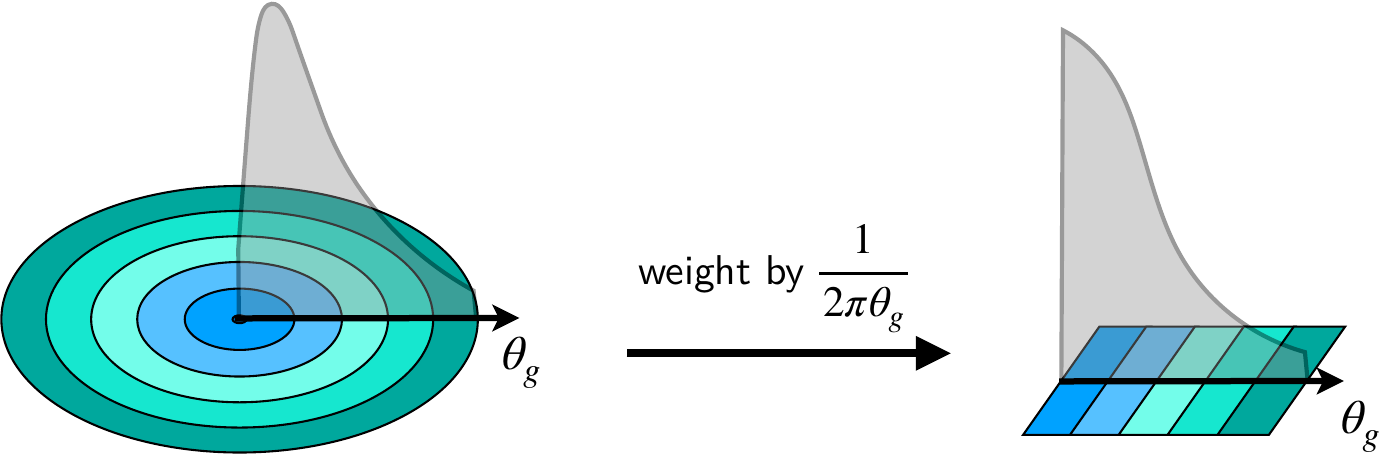}
  \caption{Transformation of the global angle $\theta_g$ to a pseudo-projected angle.}
  \label{fig:globalangle}
\end{figure*}

Multiplicative weight-based corrections are derived to correct for localised efficiency differences in the pixel detectors and geometric acceptance effects in the telescope. To minimise the geometric acceptance effects in the first place, the data-taking was divided into different segments with varied DUT positions making use of the linear stage. Residual acceptance effects are estimated by determining which fraction of tracks with an identical scatter angle would fall outside of the active area of any of the downstream planes. Over 96\% of events intersecting the measurement region on the DUT were estimated to be in the acceptance of the downstream planes.

Two types of angles were extracted from the multiplet vectors:
\begin{enumerate}
  \item{Projected angles on the global $x$-$z$ and $y$-$z$ planes, $\theta_x$ and $\theta_y$.}
  \item{Global $\phi$-invariant angles $\theta_g$, which are necessarily positive.}
\end{enumerate}

The projected angle distributions are fit with a double-sided Crystal Ball function\footnote{A double-sided Crystal Ball function is composed of a Gaussian distribution at the core, connected with two power law distributions describing the lower and upper tails.} (DSCB)~\cite{Oreglia:1980cs} in an unbinned negative log likelihood fit setup with RooFit~\cite{Verkerke:2003ir}. The fitted width of the Gaussian core of the DSCB ($\sigma_{\theta}$) is then used as an estimator for $\theta^{\text{RMS}}$. In order to treat the global angle distribution as though it were a projected distribution, each event is weighted by the inverse of the circumference of a circle with radius $\theta_g$. This procedure is sketched in Figure \ref{fig:globalangle}.

\begin{figure*}[h]
  \centering
  \subfigure[Double sided crystal ball (DSCB) fit]{\includegraphics[width=0.45\textwidth]{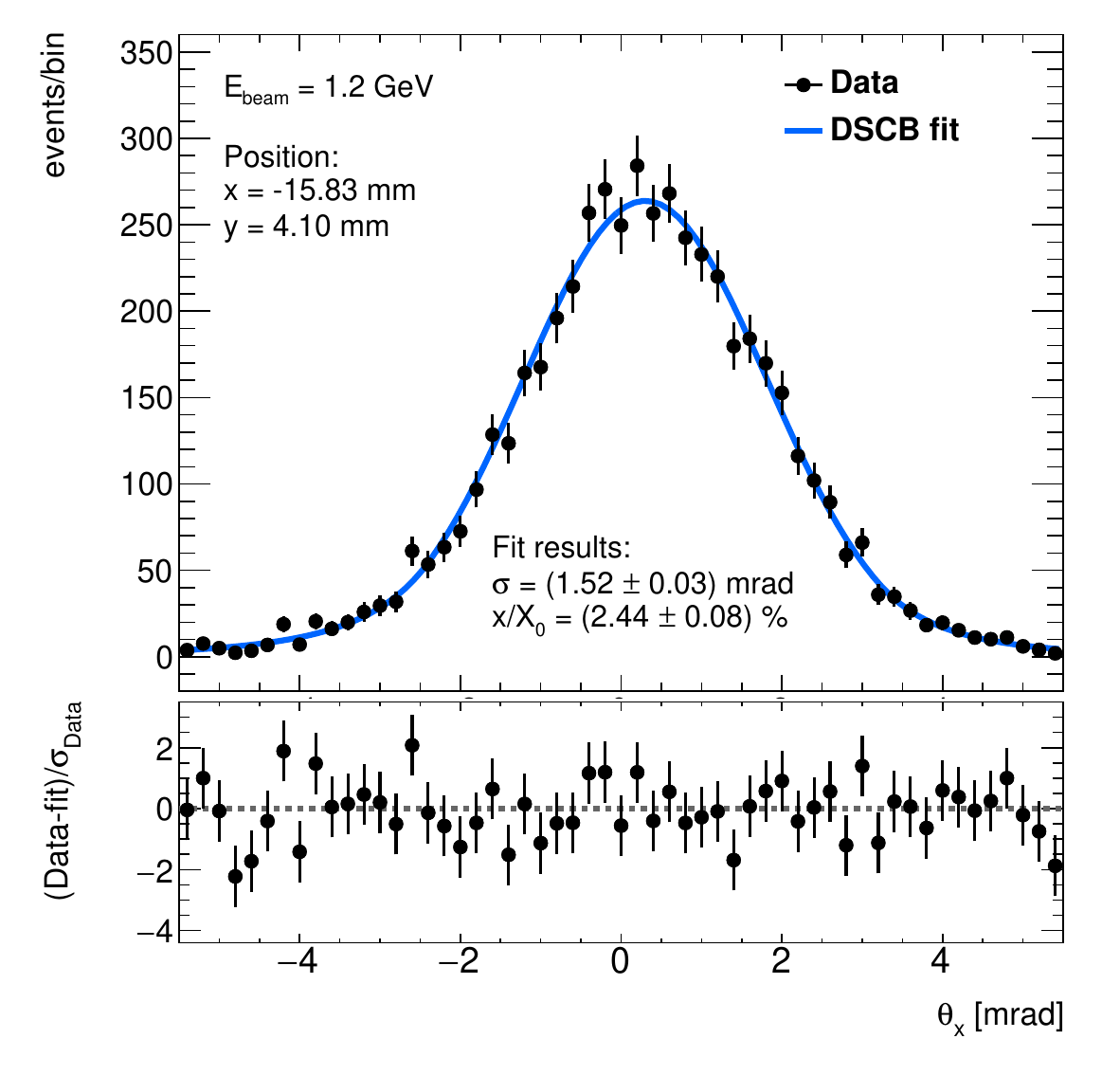}}
  \subfigure[Single sided crystal ball (SSCB) fit]{\includegraphics[width=0.45\textwidth]{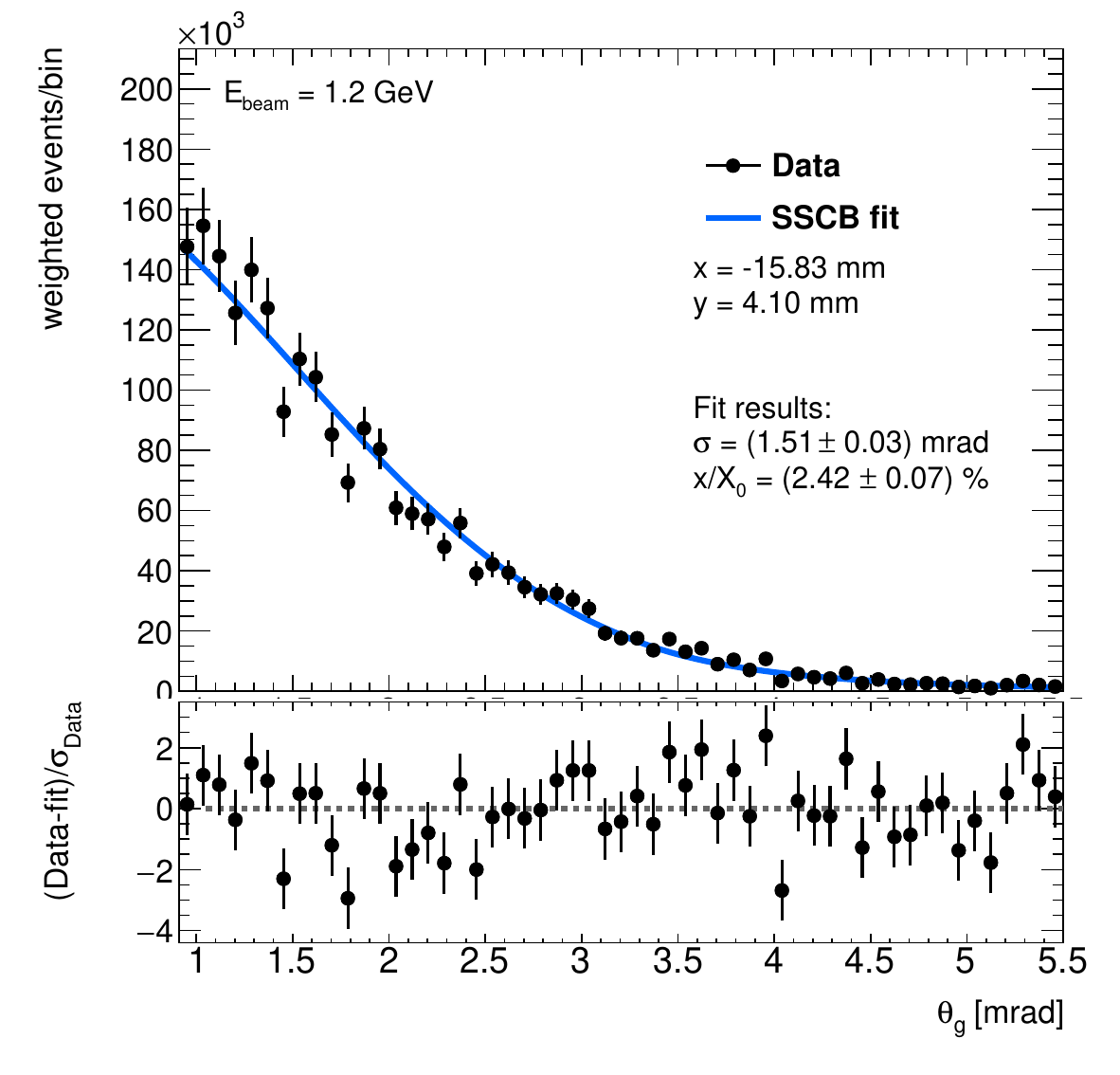}}
  \caption{Example $\theta_x$ and $\theta_g$ distributions and fit functions for one bin of one position of the DUT. The uncertainties on the quoted $x/X_0$ values include only statistical uncertainties.}
  \label{fig:fits}
\end{figure*}

The resulting distribution is then fit using a single-sided Crystal Ball function (SSCB) with the fit range adjusted to remove events on the low side due to the large uncertainties caused by events being divided by small circumferences\footnote{The lower cut is determined by averaging the $0.5 \sigma$ values from the projected angle distributions $\theta_x$ and $\theta_y$.}. Example fit results for both types of angles are shown in Figure \ref{fig:fits}.

A model of the telescope has been simulated with the Allpix$^2$ framework~\cite{Spannagel:2018usc} using the \texttt{FTFP\_BERT\_EMZ} GEANT4 physics list~\cite{GEANT4:2002zbu}, which implements the most accurate step limit for multiple scattering for the given scatter energy. Running the full reconstruction chain on the simulated scatter events, the extracted $x/X_{0}$ values are compared to the values assumed in the simulation. This allows for a comparison of the extraction based on $\theta_x$ and $\theta_y$ with the extraction using a fit to $\theta_g$, in both cases using the inverse Highland formula. All three extraction methods agree within uncertainties, both with each other and with the $x/X_{0}$ value chosen in simulation. The $\theta_x$ and $\theta_y$ based extractions have been chosen as a baseline for the results shown in the following sections. Because $\theta_x$ and $\theta_y$ are by construction orthogonal, for the final results they are combined as individual measurements. A comparison with the $\theta_g$ based extraction is shown at the end of the paper in Figure \ref{fig:full_comparison}.

\begin{figure}[h]
  \centering
  \includegraphics[width=0.45\textwidth]{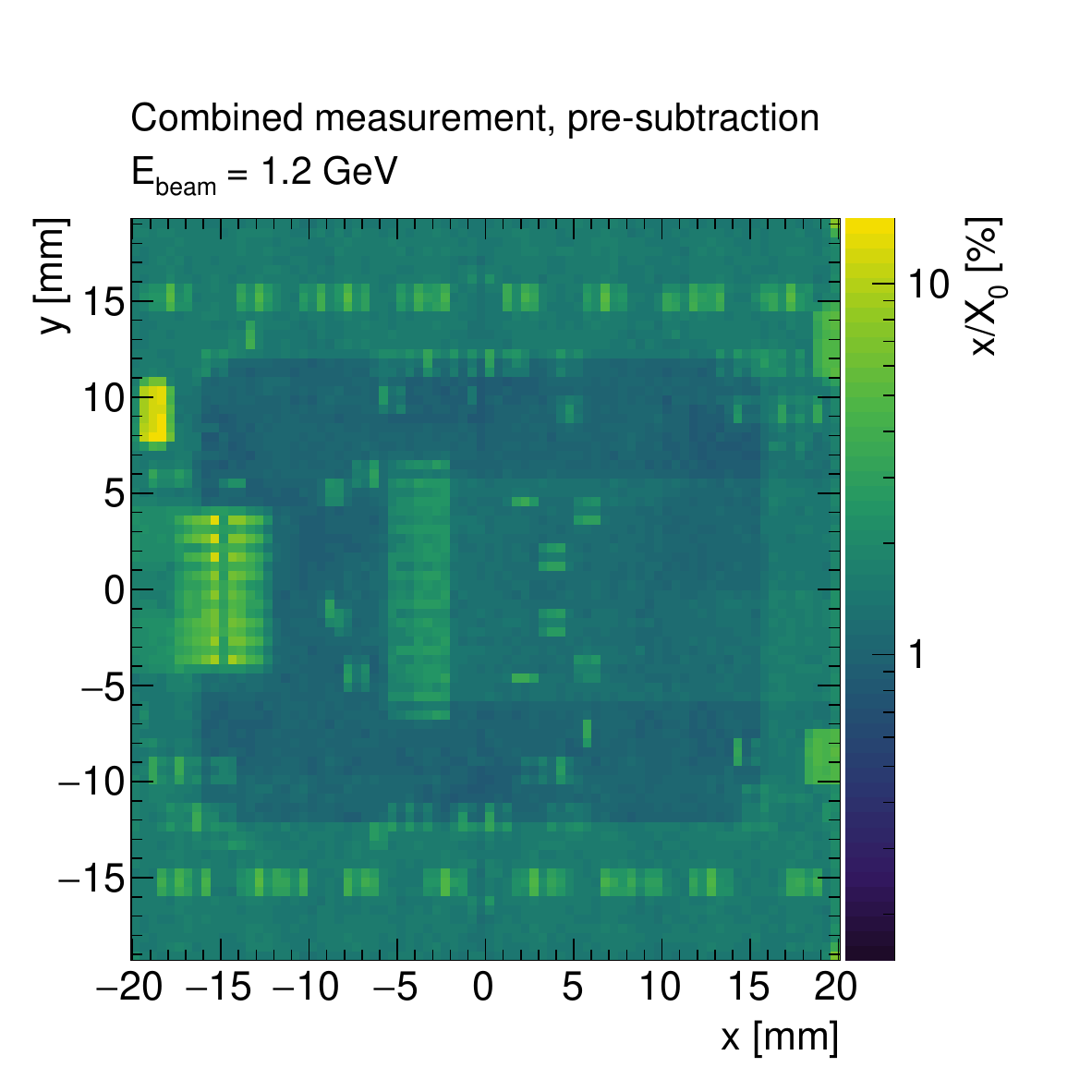}
  \caption{Measured radiation length $x/X_0$ for an ITkPix v1.1 pixel quad modue. The radiation length is extracted from the multiple scattering angle of $1.2\,\text{GeV}$ positrons in the quad module using the inverse Highland formula. The radiation length map is shown before subtraction of the air and the module holder contributions.}
  \label{fig:pre_subtraction}
\end{figure}

\begin{figure*}[h]
  \centering
  \includegraphics[width=0.85\textwidth]{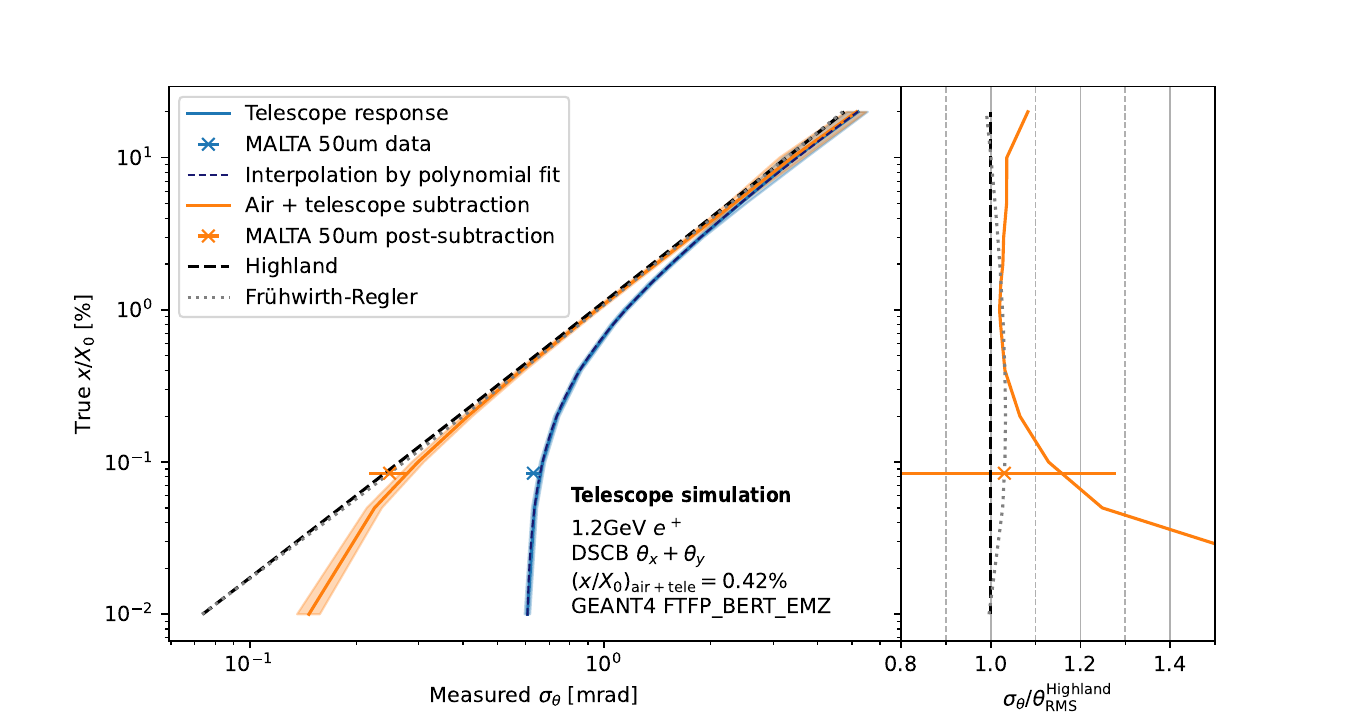}
  \caption{Simulation of the measured scatter angle RMS for various assumed true $x/X_0$ values of a homogeneous silicon DUT. The bands obtained from simulation are shown before (blue) and after the air + telescope subtraction (orange). The latter is compared to the to expectation from Highland and Fr\"uhwirth-Regler (dashed lines). Both unsubtracted and subtracted simulations are compared to the reference measurement with a $50\,\mu$m thick MALTA plane as the DUT. On the left plot of the figure, the Highland and Fr\"uhwirth-Regler lines overlap. Their difference is only visible in the ratio plot on the right of the figure.}
  \label{fig:simulation_air}
\end{figure*}

Figure \ref{fig:pre_subtraction} shows the fractional radiation length for the ITk pixel quad module under study extracted from the fitted $\theta_x$ and $\theta_y$ distributions. To arrive at the final measurement for the module, two residual corrections are made. First, the overlapping part of the aluminium frame from the module holder and the module clamp, which are both visible in Figure \ref{fig:module_holder}, need to be subtracted. Second, the scatter contribution from the telescope reference planes as well as the air between the planes needs to be accounted for. The first subtraction is made based on radiation length estimates for aluminium and polylactide\footnote{Using the analytic formula given in Ref.~\cite{Gupta:1279627} for $(\text{C}_{3}\text{H}_{4}\text{O}_{2})_{n}$.} after measuring the thickness of the components. The telescope and air contribution is estimated from simulation and cross-checked with the reference measurement of the $50\,\mu$m thick MALTA detector as DUT, which is sufficiently homogeneous. A comparison of the measurement results with the results obtained from simulation including the telescope and air contribution, as well as the agreement with the Highland and Fr\"uhwirth-Regler formulas after subtraction, is shown in Figure \ref{fig:simulation_air}. The reference measurement and the simulation agree within uncertainties, which validates the simulation. The telescope and air contribution is estimated from the measurement to $(x/X_0)_{\text{tel+air}} = (0.42 \pm 0.01)\%$. The final $x/X_{0}$ values are then calculated as

\begin{align*}
\begin{split}
        \frac{x}{X_0}(x, y) =& \left(\frac{x}{X_0}(x, y)\right)_{\text{measured}} \\&- \left(\frac{x}{X_0}(x, y)\right)_{\text{module holder}}\\ &- \left(\frac{x}{X_0}\right)_{\text{tel+air}} \quad \text{.}
\end{split}
\end{align*}

\section{Results}
\label{sec:result}

\begin{figure*}[h]
  \centering
  \subfigure[Measurement]{\includegraphics[width=0.45\textwidth]{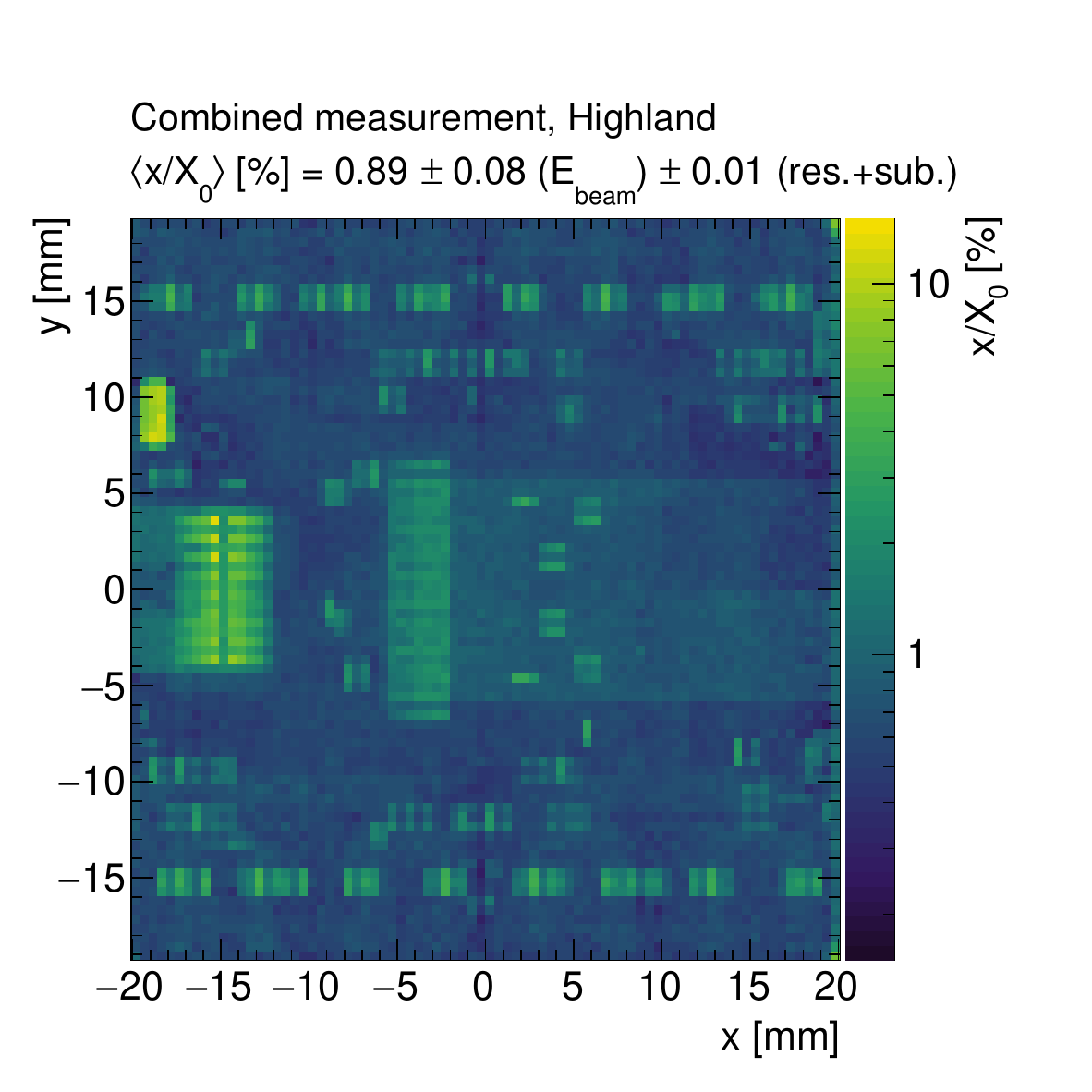}}
  \subfigure[Estimate]{\includegraphics[width=0.45\textwidth]{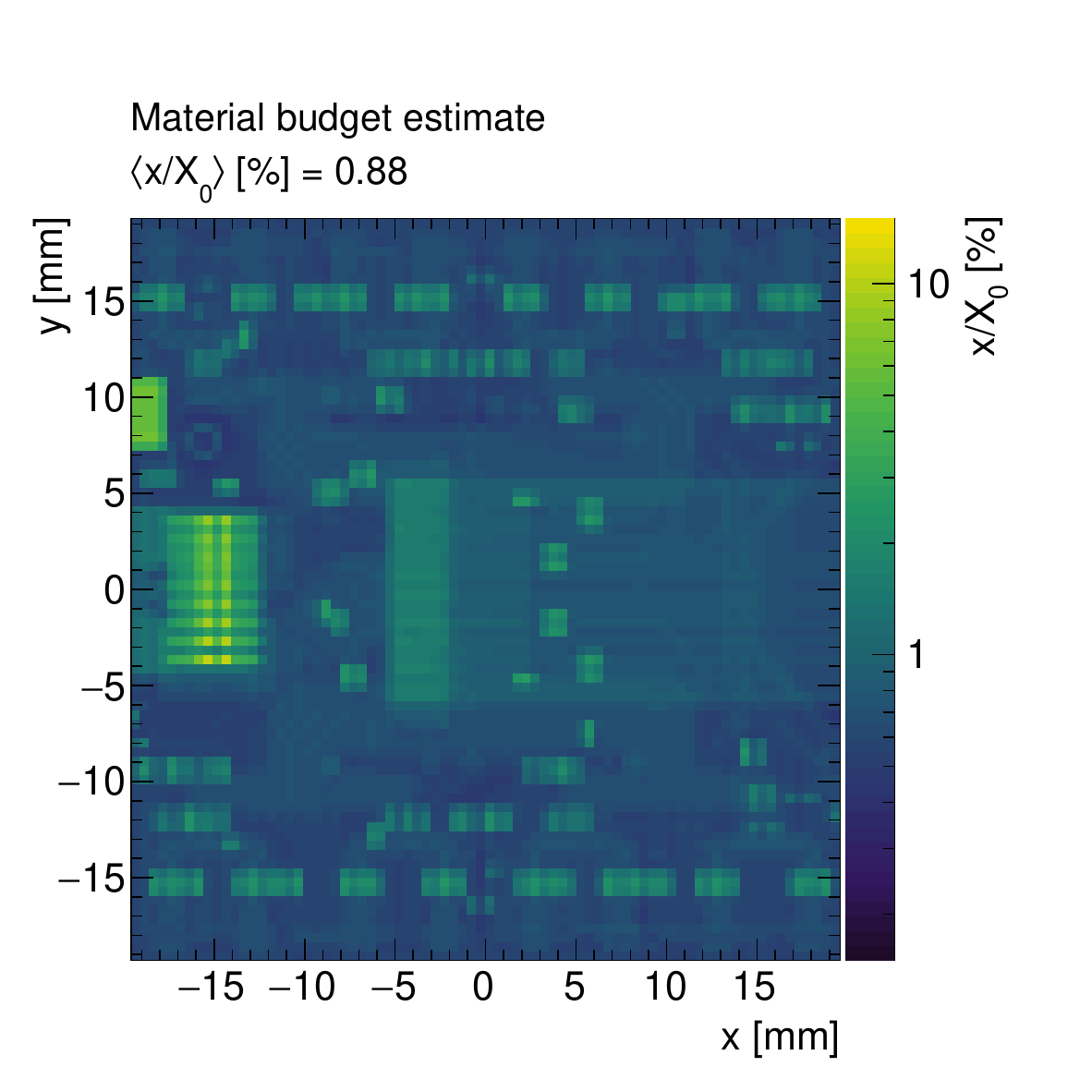}}
  \caption{Measured and estimated fractional radiation length $x/X_0$ for an ITkPix v1.1 pixel quad module, populated with a $150\,\mu\text{m}$ planar sensor and a quad flex v2.4. For the measured map, the inverse Highland formula has been used to convert the measured $\theta$ RMS values into $x/X_0$ values. The areas of largest radiation length correspond to the power and data connectors, the high voltage (HV) decoupling capacitor and the SMD components.}
  \label{fig:post_subtraction}
\end{figure*}

\begin{figure}[h]
  \centering
  \includegraphics[width=0.45\textwidth]{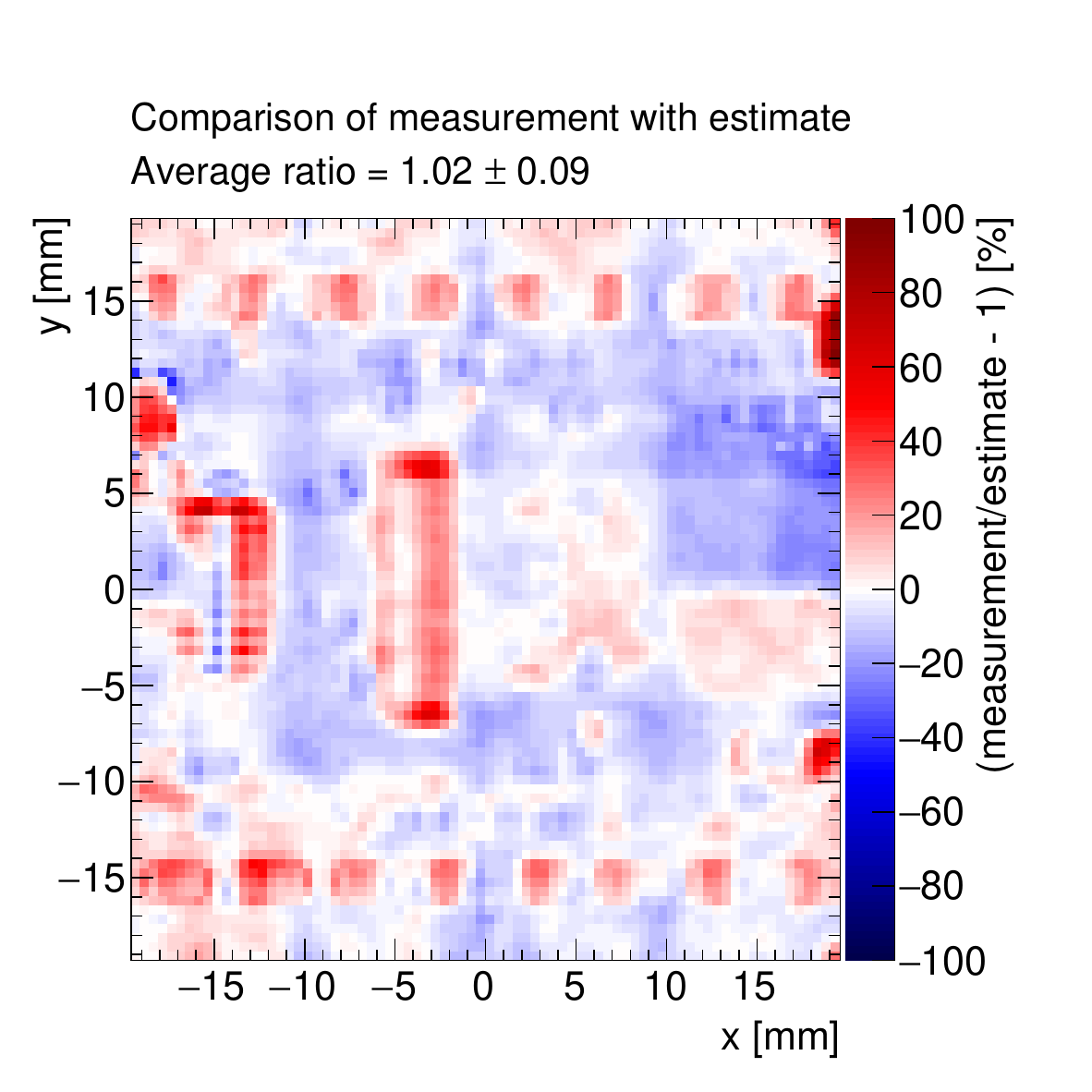}
  \caption{Comparison between the measured and expected radiation length map. The largest differences are seen at the connectors and the SMD components, related to both their material budget as well as their placement on the PCB.}
  \label{fig:comparison}
\end{figure}

The extracted $x/X_0$ map, based on the inverse Highland formula and post subtractions for the ITk pixel module under study is shown in Figure \ref{fig:post_subtraction}. The areas of largest radiation length correspond to the power and data connectors, the high voltage (HV) decoupling capacitor and the SMD components, as illustrated in Figure \ref{fig:module}. Also shown in Figure \ref{fig:post_subtraction} is an estimate of the $x/X_0$ map that has been created based on the PCB design files and product data sheets for the SMD components and connectors, where available. Where not available, best guess values have been used. The average fractional radiation length of the module is measured to

\begin{align*}
\begin{split}
  \left \langle \frac{x}{X_0} \right \rangle_{\text{meas}} [\%] = 0.89 &\pm 0.01~(\text{reso.})\\&\pm 0.01~(\text{subtraction})\\&\pm 0.08~(E_{\text{beam}}) \quad \text{,}
\end{split}
\end{align*}

which agrees with the estimate of $\langle x/X_0 \rangle_{\text{est}} = 0.88\%$. The largest source of uncertainty is the beam momentum band of $5\%$, which translates into a $9\%$ relative uncertainty on the extracted $x/X_0$ values. The statistical uncertainty on the average is negligible and about $0.05\%$ for each bin of the map presented in Figure \ref{fig:post_subtraction}. The uncertainty from the limited resolution of the telescope has been estimated by propagating a hit position uncertainty of $d/\sqrt{12}$ per plane, where $d$ is the pixel pitch, to the scatter angles. This represents a conservative estimate for the resolution of each plane. Using varied toy datasets for the scatter angle distributions, the uncertainty is determined to be $0.01\%$ and therefore subdominant. A spatially resolved comparison between estimate and measurement is shown in Figure \ref{fig:comparison}. Deviations of 50\%--100\% are seen for regions including large material contributions that may not be well-modelled in the estimate, such as proprietary connectors or surface-mount components on the module flex PCB.

\begin{figure*}[h]
  \centering
  \includegraphics[width=0.85\textwidth]{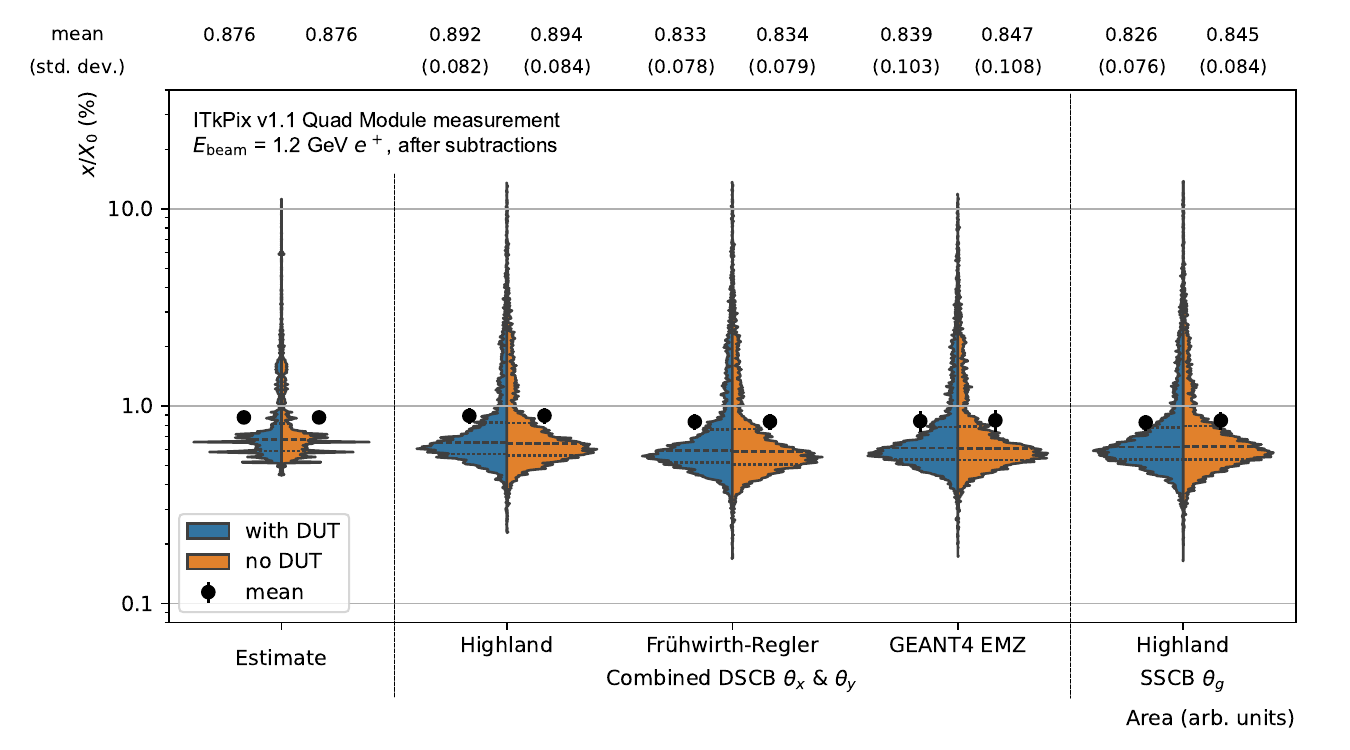}
  \caption{Violin plots comparing the estimated $x/X_0$ distribution of the ITk pixel module with three different extraction methods, based on the Highland formula, the Fr\"uhwirth-Regler formula and a GEANT4 based mapping of the measured $\theta$ to the underlying $x/X_0$ values.}
  \label{fig:full_comparison}
\end{figure*}

While the baseline results in this paper have been obtained through a combination of fits to the projected scatter angles and using the inverse Highland formula, a comparison of all three extraction methods considered is shown in Figure \ref{fig:full_comparison}. In addition to the Highland and Fr\"uhwirth-Regler formalism a third method using the GEANT4-based simulation has been performed. This method uses simulated scatters to map the fitted $\sigma_{\theta}$ to the values of $x/X_{0}$ assumed in simulation. This response is then fitted with a third-order polynomial. All three methods based on a combination of $\theta_x$ and $\theta_y$ fits agree well within uncertainties. The Highland formalism yields a $6\%-7\%$ higher average $x/X_0$ compared to Fr\"uhwirth-Regler or the GEANT4-based extraction. In addition, for the extraction based on the Highland formalism, results obtained from a fit to the global scatter angle $\theta_g$ are shown in the same figure. They also agree with the aforementioned results and the estimate within uncertainties, confirming independence of the results with respect to the fit methodology.

The $x/X_0$ values are compared between the case with including the DUT hit in the fit (blue distribution) and the case without including it (orange distribution) and agree well. For the comparison without the DUT hit, the telescope and air subtraction based on the MALTA $50\,\mu\text{m}$ reference has been repeated excluding the hit in the DUT. The subtraction without the DUT hit amounts to $(x/X_0)_{\text{tel+air}} = (0.46 \pm 0.01)\%$. The slight difference with respect to the subtraction value including the DUT hit suggests a non-trivial dependence of the air contribution on the tracking.

The estimated $x/X_{0}$ spectrum does not take into account the beam momentum spread and therefore contains large regions with identical radiation lengths. 

\section{Conclusion}
\label{sec:conclusion}

A measurement of the fractional radiation length $x/X_0$ of an ATLAS ITk pixel module using multiple scattering of positrons in a low material beam telescope has been presented. The telescope has been built for the measurement and allows variable inter-plane spacings as well as $x$-$y$ positioning of the DUT. Fitting kink angle distributions in regions of interest, an $x/X_0$ map with $\sim 0.5\,\text{mm}$ resolution and $\mathcal{O}(10\%)$ uncertainty has been derived. To calculate the $x/X_0$ values, methods based on the Highland and Fr\"uhwirth-Regler formalism, and a simulation of the telescope with GEANT4 have been utilised and compared. Within the uncertainties of the measurement, all three methods agree with each other. The Highland formalism yields an average fractional radiation length of $(0.89 \pm 0.08)\%$ which agrees with an estimate of $0.88\%$, created from design drawings and component expectations.

The fact that the results do not change significantly when removing the DUT hit information from the kink angle extraction implies that this method can be applied to non-detecting materials as well. Future iterations of the measurement would benefit from a narrower beam momentum band as well as higher positron rates, to further increase the spatial resolution.

The achieved measurement uncertainty is of the same order as the current precision to which the ATLAS inner detector material budget is known~\cite{ATLAS:2017oro}, for some components already better. The presented method could therefore be used to improve the knowledge of the material of certain components prior to physics data taking, ideally already in the R\&D phase to influence the detector design.

\bmhead{Acknowledgements}
The authors would like to acknowledge the support of the CERN staff at the T9 beamline of the Proton Synchrotron, in particular Dipanwita Banerjee for her help with the beam. The authors would further like to express their gratitude to the ATLAS ITk Pixel Collaboration for providing the ITkPix quad module. The authors would also like to thank the MALTA Collaboration for providing hardware, software, and support for the telescope used for this measurement. BM received support through Schmidt Sciences, LLC. This work was supported by the Science and Technology Facilities Council [grant numbers ST/W507726/1, ST/W000628/1]. This project has received funding from the European Union’s Horizon 2020 Research and Innovation programme under Grant Agreement numbers 101004761 (AIDAinnova), 675587 (STREAM), and 654168 (IJS, Ljubljana, Slovenia).

\begingroup
\let\emph\relax
\let\textbf\relax
\bibliography{bibliography.bib}


\begin{thebibliography}{27}
\ifx \bisbn   \undefined \def \bisbn  #1{ISBN #1}\fi
\ifx \binits  \undefined \def \binits#1{#1}\fi
\ifx \bauthor  \undefined \def \bauthor#1{#1}\fi
\ifx \batitle  \undefined \def \batitle#1{#1}\fi
\ifx \bjtitle  \undefined \def \bjtitle#1{#1}\fi
\ifx \bvolume  \undefined \def \bvolume#1{\textbf{#1}}\fi
\ifx \byear  \undefined \def \byear#1{#1}\fi
\ifx \bissue  \undefined \def \bissue#1{#1}\fi
\ifx \bfpage  \undefined \def \bfpage#1{#1}\fi
\ifx \blpage  \undefined \def \blpage #1{#1}\fi
\ifx \burl  \undefined \def \burl#1{\textsf{#1}}\fi
\ifx \doiurl  \undefined \def \doiurl#1{\url{https://doi.org/#1}}\fi
\ifx \betal  \undefined \def \betal{\textit{et al.}}\fi
\ifx \binstitute  \undefined \def \binstitute#1{#1}\fi
\ifx \binstitutionaled  \undefined \def \binstitutionaled#1{#1}\fi
\ifx \bctitle  \undefined \def \bctitle#1{#1}\fi
\ifx \beditor  \undefined \def \beditor#1{#1}\fi
\ifx \bpublisher  \undefined \def \bpublisher#1{#1}\fi
\ifx \bbtitle  \undefined \def \bbtitle#1{#1}\fi
\ifx \bedition  \undefined \def \bedition#1{#1}\fi
\ifx \bseriesno  \undefined \def \bseriesno#1{#1}\fi
\ifx \blocation  \undefined \def \blocation#1{#1}\fi
\ifx \bsertitle  \undefined \def \bsertitle#1{#1}\fi
\ifx \bsnm \undefined \def \bsnm#1{#1}\fi
\ifx \bsuffix \undefined \def \bsuffix#1{#1}\fi
\ifx \bparticle \undefined \def \bparticle#1{#1}\fi
\ifx \barticle \undefined \def \barticle#1{#1}\fi
\bibcommenthead
\ifx \bconfdate \undefined \def \bconfdate #1{#1}\fi
\ifx \botherref \undefined \def \botherref #1{#1}\fi
\ifx \url \undefined \def \url#1{\textsf{#1}}\fi
\ifx \bchapter \undefined \def \bchapter#1{#1}\fi
\ifx \bbook \undefined \def \bbook#1{#1}\fi
\ifx \bcomment \undefined \def \bcomment#1{#1}\fi
\ifx \oauthor \undefined \def \oauthor#1{#1}\fi
\ifx \citeauthoryear \undefined \def \citeauthoryear#1{#1}\fi
\ifx \endbibitem  \undefined \def \endbibitem {}\fi
\ifx \bconflocation  \undefined \def \bconflocation#1{#1}\fi
\ifx \arxivurl  \undefined \def \arxivurl#1{\textsf{#1}}\fi
\csname PreBibitemsHook\endcsname

\bibitem[\protect\citeauthoryear{Zurbano~Fernandez et~al.}{2020}]{ZurbanoFernandez:2020cco}
\begin{botherref}
\oauthor{\bsnm{Zurbano~Fernandez}, \binits{I.}}, et al.:
{High-Luminosity Large Hadron Collider (HL-LHC): Technical design report}
\textbf{10/2020}
(2020)
\doiurl{10.23731/CYRM-2020-0010}
\end{botherref}
\endbibitem

\bibitem[\protect\citeauthoryear{{ATLAS Collaboration}}{2017a}]{CERN-LHCC-2017-005}
\begin{botherref}
\oauthor{\bsnm{{ATLAS Collaboration}}}:
{Technical Design Report for the ATLAS Inner Tracker Strip Detector}.
Technical report,
CERN,
Geneva
(2017).
\url{https://cds.cern.ch/record/2257755}
\end{botherref}
\endbibitem

\bibitem[\protect\citeauthoryear{{ATLAS Collaboration}}{2017b}]{CERN-LHCC-2017-021}
\begin{botherref}
\oauthor{\bsnm{{ATLAS Collaboration}}}:
{Technical Design Report for the ATLAS Inner Tracker Pixel Detector}.
Technical report,
CERN,
Geneva
(2017).
\doiurl{10.17181/CERN.FOZZ.ZP3Q} .
\url{https://cds.cern.ch/record/2285585}
\end{botherref}
\endbibitem

\bibitem[\protect\citeauthoryear{{ATLAS Collaboration}}{2019}]{ATL-PHYS-PUB-2019-014}
\begin{botherref}
\oauthor{\bsnm{{ATLAS Collaboration}}}:
{Expected Tracking Performance of the ATLAS Inner Tracker at the HL-LHC}.
Technical report,
CERN,
Geneva
(2019).
\url{https://cds.cern.ch/record/2669540}
\end{botherref}
\endbibitem

\bibitem[\protect\citeauthoryear{Garcia-Sciveres et~al.}{2019}]{Garcia-Sciveres:2665301}
\begin{botherref}
\oauthor{\bsnm{Garcia-Sciveres}, \binits{M.}},
\oauthor{\bsnm{Loddo}, \binits{F.}},
\oauthor{\bsnm{Christiansen}, \binits{J.}}:
{RD53B Manual}.
Technical report,
CERN,
Geneva
(2019).
\url{https://cds.cern.ch/record/2665301}
\end{botherref}
\endbibitem

\bibitem[\protect\citeauthoryear{Bethe}{1953}]{PhysRev.89.1256}
\begin{barticle}
\bauthor{\bsnm{Bethe}, \binits{H.A.}}:
\batitle{Moli\`ere's theory of multiple scattering}.
\bjtitle{Phys. Rev.}
\bvolume{89},
\bfpage{1256}--\blpage{1266}
(\byear{1953})
\doiurl{10.1103/PhysRev.89.1256}
\end{barticle}
\endbibitem

\bibitem[\protect\citeauthoryear{Lynch and Dahl}{1991}]{Lynch:1990sq}
\begin{barticle}
\bauthor{\bsnm{Lynch}, \binits{G.R.}},
\bauthor{\bsnm{Dahl}, \binits{O.I.}}:
\batitle{{Approximations to multiple Coulomb scattering}}.
\bjtitle{Nucl. Instrum. Meth. B}
\bvolume{58},
\bfpage{6}--\blpage{10}
(\byear{1991})
\doiurl{10.1016/0168-583X(91)95671-Y}
\end{barticle}
\endbibitem

\bibitem[\protect\citeauthoryear{Group}{2022}]{highlandpdg}
\begin{barticle}
\bauthor{\bsnm{Group}, \binits{P.D.}}:
\batitle{{Review of Particle Physics}}.
\bjtitle{Progress of Theoretical and Experimental Physics}
\bvolume{2022}(\bissue{8}),
\bfpage{083}--\blpage{01}
(\byear{2022})
\doiurl{10.1093/ptep/ptac097}
{\href{https://arxiv.org/abs/https://academic.oup.com/ptep/article-pdf/2022/8/083C01/49175539/ptac097.pdf}{{https://academic.oup.com/ptep/article-pdf/2022/8/083C01/49175539/ptac097.pdf}}}
\end{barticle}
\endbibitem

\bibitem[\protect\citeauthoryear{Frühwirth and Regler}{2001}]{FRUHWIRTH2001369}
\begin{barticle}
\bauthor{\bsnm{Frühwirth}, \binits{R.}},
\bauthor{\bsnm{Regler}, \binits{M.}}:
\batitle{On the quantitative modelling of core and tails of multiple scattering by gaussian mixtures}.
\bjtitle{Nuclear Instruments and Methods in Physics Research Section A: Accelerators, Spectrometers, Detectors and Associated Equipment}
\bvolume{456}(\bissue{3}),
\bfpage{369}--\blpage{389}
(\byear{2001})
\doiurl{10.1016/S0168-9002(00)00589-1}
\end{barticle}
\endbibitem

\bibitem[\protect\citeauthoryear{Stolzenberg et~al.}{2017}]{STOLZENBERG2017173}
\begin{barticle}
\bauthor{\bsnm{Stolzenberg}, \binits{U.}},
\bauthor{\bsnm{Frey}, \binits{A.}},
\bauthor{\bsnm{Schwenker}, \binits{B.}},
\bauthor{\bsnm{Wieduwilt}, \binits{P.}},
\bauthor{\bsnm{Marinas}, \binits{C.}},
\bauthor{\bsnm{Lütticke}, \binits{F.}}:
\batitle{Radiation length imaging with high-resolution telescopes}.
\bjtitle{Nuclear Instruments and Methods in Physics Research Section A: Accelerators, Spectrometers, Detectors and Associated Equipment}
\bvolume{845},
\bfpage{173}--\blpage{176}
(\byear{2017})
\doiurl{10.1016/j.nima.2016.06.086} .
\bcomment{Proceedings of the Vienna Conference on Instrumentation 2016}
\end{barticle}
\endbibitem

\bibitem[\protect\citeauthoryear{Poley et~al.}{2021}]{Poley:2021wcw}
\begin{barticle}
\bauthor{\bsnm{Poley}, \binits{L.}},
\bauthor{\bsnm{Stolzenberg}, \binits{U.}},
\bauthor{\bsnm{Schwenker}, \binits{B.}},
\bauthor{\bsnm{Frey}, \binits{A.}},
\bauthor{\bsnm{G\"ottlicher}, \binits{P.}},
\bauthor{\bsnm{Marinas}, \binits{C.}},
\bauthor{\bsnm{Stanitzki}, \binits{M.}},
\bauthor{\bsnm{Stelzer}, \binits{B.}}:
\batitle{{Mapping the material distribution of a complex structure in an electron beam}}.
\bjtitle{JINST}
\bvolume{16}(\bissue{01}),
\bfpage{01010}
(\byear{2021})
\doiurl{10.1088/1748-0221/16/01/P01010}
{\href{https://arxiv.org/abs/2105.10128}{{arXiv:2105.10128}}}
{[physics.ins-det]}
\end{barticle}
\endbibitem

\bibitem[\protect\citeauthoryear{Qu et~al.}{2021}]{Qu:2021qda}
\begin{barticle}
\bauthor{\bsnm{Qu}, \binits{C.Y.}}, \betal:
\batitle{{Analysis of pixel detector material budget based on test beam}}.
\bjtitle{JINST}
\bvolume{16}(\bissue{06}),
\bfpage{06004}
(\byear{2021})
\doiurl{10.1088/1748-0221/16/06/T06004}
\end{barticle}
\endbibitem

\bibitem[\protect\citeauthoryear{{Tracker Group of the CMS Collaboration}}{2024}]{TrackerGroupoftheCMS:2024utw}
\begin{barticle}
\bauthor{\bsnm{{Tracker Group of the CMS Collaboration}}}:
\batitle{{Measurement of the fractional radiation length of a pixel module for the CMS Phase-2 upgrade via the multiple scattering of positrons}}.
\bjtitle{JINST}
\bvolume{19}(\bissue{10}),
\bfpage{10023}
(\byear{2024})
\doiurl{10.1088/1748-0221/19/10/P10023}
{\href{https://arxiv.org/abs/2407.13721}{{arXiv:2407.13721}}}
{[physics.ins-det]}
\end{barticle}
\endbibitem

\bibitem[\protect\citeauthoryear{Solans~S\'anchez et~al.}{2023}]{SolansSanchez:2023pik}
\begin{barticle}
\bauthor{\bsnm{Solans~S\'anchez}, \binits{C.}}, \betal:
\batitle{{MALTA monolithic pixel sensors in TowerJazz 180 nm technology}}.
\bjtitle{Nucl. Instrum. Meth. A}
\bvolume{1057},
\bfpage{168787}
(\byear{2023})
\doiurl{10.1016/j.nima.2023.168787}
\end{barticle}
\endbibitem

\bibitem[\protect\citeauthoryear{Pernegger et~al.}{2023}]{Pernegger:2023yaf}
\begin{barticle}
\bauthor{\bsnm{Pernegger}, \binits{H.}}, \betal:
\batitle{{MALTA-Cz: a radiation hard full-size monolithic CMOS sensor with small electrodes on high-resistivity Czochralski substrate}}.
\bjtitle{JINST}
\bvolume{18}(\bissue{09}),
\bfpage{09018}
(\byear{2023})
\doiurl{10.1088/1748-0221/18/09/P09018}
{\href{https://arxiv.org/abs/2301.03912}{{arXiv:2301.03912}}}
{[physics.ins-det]}
\end{barticle}
\endbibitem

\bibitem[\protect\citeauthoryear{van Rijnbach et~al.}{2023}]{vanRijnbach:2023qgn}
\begin{barticle}
\bauthor{\bsnm{Rijnbach}, \binits{M.}}, \betal:
\batitle{{Performance of the MALTA telescope}}.
\bjtitle{Eur. Phys. J. C}
\bvolume{83}(\bissue{7}),
\bfpage{581}
(\byear{2023})
\doiurl{10.1140/epjc/s10052-023-11760-z}
{\href{https://arxiv.org/abs/2304.01104}{{arXiv:2304.01104}}}
{[hep-ex]}
\end{barticle}
\endbibitem

\bibitem[\protect\citeauthoryear{Franconi and on~behalf~of ATLAS~ITk}{2022}]{Franconi_2022}
\begin{barticle}
\bauthor{\bsnm{Franconi}, \binits{L.}},
\bauthor{\bsnm{ATLAS~ITk}}:
\batitle{The opto-electrical conversion system for the data transmission chain of the atlas itk pixel detector upgrade for the hl-lhc}.
\bjtitle{Journal of Physics: Conference Series}
\bvolume{2374}(\bissue{1}),
\bfpage{012105}
(\byear{2022})
\doiurl{10.1088/1742-6596/2374/1/012105}
\end{barticle}
\endbibitem

\bibitem[\protect\citeauthoryear{Ryu and on~behalf of~the ATLAS TDAQ~Collaboration}{2017}]{Ryu_2017}
\begin{barticle}
\bauthor{\bsnm{Ryu}, \binits{S.}},
\bauthor{\bsnm{ATLAS TDAQ~Collaboration}}:
\batitle{Felix: The new detector readout system for the atlas experiment}.
\bjtitle{Journal of Physics: Conference Series}
\bvolume{898}(\bissue{3}),
\bfpage{032057}
(\byear{2017})
\doiurl{10.1088/1742-6596/898/3/032057}
\end{barticle}
\endbibitem

\bibitem[\protect\citeauthoryear{Sanchez and on~behalf of~the ATLAS~Collaboration}{2020}]{Solans_Sanchez_2020}
\begin{barticle}
\bauthor{\bsnm{Sanchez}, \binits{C.A.S.}},
\bauthor{\bsnm{ATLAS~Collaboration}}:
\batitle{The felix detector interface for the atlas tdaq upgrades and its deployment in the itk demonstrator setup}.
\bjtitle{Journal of Physics: Conference Series}
\bvolume{1525}(\bissue{1}),
\bfpage{012033}
(\byear{2020})
\doiurl{10.1088/1742-6596/1525/1/012033}
\end{barticle}
\endbibitem

\bibitem[\protect\citeauthoryear{Dannheim et~al.}{2021}]{Dannheim_2021}
\begin{barticle}
\bauthor{\bsnm{Dannheim}, \binits{D.}},
\bauthor{\bsnm{Dort}, \binits{K.}},
\bauthor{\bsnm{Huth}, \binits{L.}},
\bauthor{\bsnm{Hynds}, \binits{D.}},
\bauthor{\bsnm{Kremastiotis}, \binits{I.}},
\bauthor{\bsnm{Kröger}, \binits{J.}},
\bauthor{\bsnm{Munker}, \binits{M.}},
\bauthor{\bsnm{Pitters}, \binits{F.}},
\bauthor{\bsnm{Schütze}, \binits{P.}},
\bauthor{\bsnm{Spannagel}, \binits{S.}},
\bauthor{\bsnm{Vanat}, \binits{T.}},
\bauthor{\bsnm{Williams}, \binits{M.}}:
\batitle{Corryvreckan: a modular 4d track reconstruction and analysis software for test beam data}.
\bjtitle{Journal of Instrumentation}
\bvolume{16}(\bissue{03}),
\bfpage{03008}
(\byear{2021})
\doiurl{10.1088/1748-0221/16/03/P03008}
\end{barticle}
\endbibitem

\bibitem[\protect\citeauthoryear{Blobel and Kleinwort}{2002}]{Blobel:2002ax}
\begin{bchapter}
\bauthor{\bsnm{Blobel}, \binits{V.}},
\bauthor{\bsnm{Kleinwort}, \binits{C.}}:
\bctitle{{A New method for the high precision alignment of track detectors}}.
In: \bbtitle{{Conference on Advanced Statistical Techniques in Particle Physics}}
(\byear{2002})
\end{bchapter}
\endbibitem

\bibitem[\protect\citeauthoryear{Oreglia}{1980}]{Oreglia:1980cs}
\begin{botherref}
\oauthor{\bsnm{Oreglia}, \binits{M.}}:
{A Study of the Reactions $\psi^\prime \to \gamma \gamma \psi$}.
Phd. thesis
(December 1980).
\url{www.slac.stanford.edu/cgi-wrap/getdoc/slac-r-236.pdf}
\end{botherref}
\endbibitem

\bibitem[\protect\citeauthoryear{Verkerke and Kirkby}{2003}]{Verkerke:2003ir}
\begin{barticle}
\bauthor{\bsnm{Verkerke}, \binits{W.}},
\bauthor{\bsnm{Kirkby}, \binits{D.P.}}:
\batitle{{The RooFit toolkit for data modeling}}.
\bjtitle{eConf}
\bvolume{C0303241},
\bfpage{007}
(\byear{2003})
{\href{https://arxiv.org/abs/physics/0306116}{{arXiv:physics/0306116}}}
\end{barticle}
\endbibitem

\bibitem[\protect\citeauthoryear{Spannagel et~al.}{2018}]{Spannagel:2018usc}
\begin{barticle}
\bauthor{\bsnm{Spannagel}, \binits{S.}},
\bauthor{\bsnm{Wolters}, \binits{K.}},
\bauthor{\bsnm{Hynds}, \binits{D.}},
\bauthor{\bsnm{Alipour~Tehrani}, \binits{N.}},
\bauthor{\bsnm{Benoit}, \binits{M.}},
\bauthor{\bsnm{Dannheim}, \binits{D.}},
\bauthor{\bsnm{Gauvin}, \binits{N.}},
\bauthor{\bsnm{N\"urnberg}, \binits{A.}},
\bauthor{\bsnm{Sch\"utze}, \binits{P.}},
\bauthor{\bsnm{Vicente Barreto~Pinto}, \binits{M.}}:
\batitle{{Allpix$^2$: A Modular Simulation Framework for Silicon Detectors}}.
\bjtitle{Nucl. Instrum. Meth. A}
\bvolume{901},
\bfpage{164}--\blpage{172}
(\byear{2018})
\doiurl{10.1016/j.nima.2018.06.020}
{\href{https://arxiv.org/abs/1806.05813}{{arXiv:1806.05813}}}
{[physics.ins-det]}
\end{barticle}
\endbibitem

\bibitem[\protect\citeauthoryear{Agostinelli et~al.}{2003}]{GEANT4:2002zbu}
\begin{barticle}
\bauthor{\bsnm{Agostinelli}, \binits{S.}}, \betal:
\batitle{{GEANT4--a simulation toolkit}}.
\bjtitle{Nucl. Instrum. Meth. A}
\bvolume{506},
\bfpage{250}--\blpage{303}
(\byear{2003})
\doiurl{10.1016/S0168-9002(03)01368-8}
\end{barticle}
\endbibitem

\bibitem[\protect\citeauthoryear{Gupta}{2010}]{Gupta:1279627}
\begin{botherref}
\oauthor{\bsnm{Gupta}, \binits{M.}}:
{Calculation of radiation length in materials}.
Technical report,
CERN,
Geneva
(2010).
\url{https://cds.cern.ch/record/1279627}
\end{botherref}
\endbibitem

\bibitem[\protect\citeauthoryear{{ATLAS Collaboration}}{2017}]{ATLAS:2017oro}
\begin{barticle}
\bauthor{\bsnm{{ATLAS Collaboration}}}:
\batitle{{Study of the material of the ATLAS inner detector for Run 2 of the LHC}}.
\bjtitle{JINST}
\bvolume{12}(\bissue{12}),
\bfpage{12009}
(\byear{2017})
\doiurl{10.1088/1748-0221/12/12/P12009}
{\href{https://arxiv.org/abs/1707.02826}{{arXiv:1707.02826}}}
{[hep-ex]}
\end{barticle}
\endbibitem

\end{thebibliography}
\endgroup

\end{document}